\documentclass[a4paper,11pt]{article}
\pdfoutput=1

\usepackage{jheppub}

\usepackage[T1]{fontenc}
\usepackage[utf8]{inputenc}

\usepackage{amsmath,amssymb,braket}
\usepackage{pifont}%
\usepackage{booktabs,multirow,boldline}
\usepackage{enumitem}

\usepackage{color}
\usepackage{array}
\usepackage{graphicx}
\usepackage[dvipsnames]{xcolor}
\usepackage[normalem]{ulem}

\newcommand{\eqnref}[1]{eq.~(\ref{#1})}

 \newcommand{\eqn}[1]{eq.~(\ref{#1})}
 \newcommand{\eqns}[2]{eqs.~(\ref{#1})-(\ref{#2})}

\newcommand{\mX}{\mathcal{X}}

\newcommand{\cmark}{\ding{51}}%
\newcommand{\xmark}{\ding{55}}%

\definecolor{nicegreen}{rgb}{0.1,0.5,0.1}

\newcommand{\overbar}[1]{\mkern 1.5mu\overline{\mkern-1.5mu#1\mkern-1.5mu}\mkern 1.5mu}
\newcommand{\pmatr}[1]{\begin{pmatrix} #1 \end{pmatrix}}

\DeclareMathOperator{\diag}{diag}
\DeclareMathOperator{\tr}{tr}

\newcommand{\impart}[1]{\mathrm{Im}\left[#1\right]}

\newcommand{\strucA}{\mathbf{1}}
\newcommand{\strucB}{\mathbf{0}}

\newcommand{\mfb}{m}

\newcommand{\UVX}[1]{\mathbf{#1}}

\newcommand{\cond}{\mathcal{C}}

\newcommand{\ugd}{U(1)}

\newcommand{\ppmatr}[1]{{
	\setlength\arraycolsep{1pt}
	\renewcommand*{\arraystretch}{.8}
	\scriptstyle
	\pmatr{#1}
}}

 \newcommand{\MNxx}{$\ppmatr{\times&0&0\\0&0&\times\\0&\times&0}$}
 \newcommand{\MNxxx}{$\ppmatr{\times&\times&0\\\times&0&\times\\0&\times&0}$}

\newcommand{\pmuptw}{$\ppmatr{\times&\times&\zro\\\zro&\times&\zro\\\zro&\zro&\times}$}
\newcommand{\pmdotw}{$\ppmatr{\times&\zro&\zro\\\times&\times&\zro\\\zro&\zro&\times}$}

\newcommand{\zro}{0}

\newcommand{\NoneNone}{ $\ppmatr{ \times &\zro   &\zro   \\ \zro   &\times &\zro   \\ \zro   &\zro   &\times }$}

\newcommand{\OneTwoSym}{   $\ppmatr{ \times &\times &\zro   \\ \times &\times &\zro   \\ \zro   &\zro   &\times }$}
\newcommand{\OneThreeSym}{ $\ppmatr{ \times &\zro   &\times \\ \zro   &\times &\zro   \\ \times &\zro   &\times }$}
\newcommand{\TwoThreeSym}{ $\ppmatr{ \times &\zro   &\zro   \\ \zro   &\times &\times \\ \zro   &\times &\times }$}

\newcommand{\mnuRibbon}{$\ppmatr{ \times &\times &\zro   \\ \times &\times &\times \\ \zro   &\times &\zro   }$}
\newcommand{\mnuIsland}{$\ppmatr{ \times &\zro   &\times \\ \zro   &\zro   &\times \\ \times &\times &\times }$}
\newcommand{\mnuFarIsland}{$\ppmatr{ \times &\zro   &\times \\ \zro   &\zro   &\times \\ \times &\times &\times }$}
\newcommand{\mnuArchipelago}{$\ppmatr{ \times &\zro   &\times \\ \zro   &\zro   &\times \\ \times &\times &\zro }$}

\newcommand{\mnuOneTZLower}{$\ppmatr{\times&\times&\times\\\times&\times&\times\\\times&\times&\zro}$}
\newcommand{\mnuOneTZMiddle}{$\ppmatr{\times&\times&\times\\\times&\zro&\times\\\times&\times&\times}$}
\newcommand{\mnuFull}{$\ppmatr{\times&\times&\times\\\times&\times&\times\\\times&\times&\times}$}

\title{
Covert  
symmetries in  the neutrino mass matrix
}

\author[a]{Fredrik Bj\"orkeroth,}
\author[b,c]{Luca Di Luzio,}
\author[d]{Federico Mescia,}
\author[a]{Enrico Nardi}

\emailAdd{fredrik.bjorkeroth@gmail.com} 
\emailAdd{luca.diluzio@desy.de}
\emailAdd{mescia@ub.edu}
\emailAdd{enrico.nardi@lnf.infn.it}

\affiliation[a]{INFN, Laboratori Nazionali di Frascati, C.P. 13, 100044 Frascati, Italy}
\affiliation[b]{%
  Deutsches Elektronen-Synchrotron DESY, Notkestra\ss e 85, D-22607 Hamburg, Germany
} 
\affiliation[c]{%
  Dipartimento di Fisica, Universit\`a di Pisa and INFN, Sezione di Pisa, \\ 
  Largo B.~Pontecorvo 3, 56127 Pisa, Italy
} 
\affiliation[d]{%
  Departament de F\'isica Qu\`antica i Astrof\'isica, Institut de Ci\`encies del Cosmos (ICCUB), \\
  Universitat de Barcelona, Mart\'i Franqu\`es 1, E08028 Barcelona, Spain
}

\abstract{%
The flavour neutrino puzzle is often addressed by considering neutrino mass matrices $m$
 with a certain number of vanishing entries ($m_{ij}=0$ for some values of the indices),
since a reduction in  the number of free parameters increases the  predictive power. Symmetries 
that can enforce textures zero can also  enforce a more general type of conditions
$f(m_{ij})=0$ with $f$ some function of the matrix elements  $m_{ij}$. In this case  $m$ 
can have  all entries non-vanishing with no reduction in its predictive power. 
We  classify all  generation-dependent $U(1)$ symmetries which, in the presence of two leptonic Higgs doublets, 
can reduce the number of  independent high-energy  parameters of type-I seesaw to the minimum number 
compatible with non-vanishing neutrino mixings and CP violation.
These symmetries are broken above the scale where the effective operator is generated and   
can thus remain covert, in the sense that no explicit  evidence of the symmetry can be read off the 
neutrino mass matrix, and  different symmetries can give rise  to the same low-energy structure.
We find that only two  cases are viable:  one yields a  
  structure with  two zero-textures already considered in the literature, the other has no zero-textures and 
  has never been considered before. It predicts normal ordering, a lightest neutrino mass $ \sim 10$ meV, a Dirac phase 
  $\delta \sim  \frac{3\pi}{2}$ and  definite values for the Majorana phases.
} 

\keywords{}
\preprint{DESY 20-008}

\begin{document}
\maketitle
\flushbottom

\section{Introduction}

Flavour symmetry groups have been  widely used in 
attempts to account for the mass and mixing patterns 
of the Standard Model (SM) quarks and leptons.
Fermions are often assumed to transform in representations of 
elementary continuous groups like 
$U(1)$~\cite{
Froggatt:1978nt,
Leurer:1992wg,
Leurer:1993gy,
Dine:1993np,
Ibanez:1994ig,
Banks:1995by, 
Dudas:1996fe,
Irges:1998ax,
Mira:1999fx,
Elwood:1998kf},
$U(2)$~\cite{
Pomarol:1995xc,
Barbieri:1995uv,
Barbieri:1996ww,
Barbieri:1997tu,
Carone:1997qg}, 
$SU(3)$~\cite{
Nardi:2011st,
Alonso:2011yg,
Espinosa:2012uu,
Fong:2013dnk},  
spontaneously broken by the vacuum expectation values (vevs) of scalar
fields which are generally singlets of the SM or in the adjoint of
some GUT group~\cite{
Duque:2008ah,
Wang:2011ub,
Nardi:2011jp}.
%
%
The symmetry and symmetry-breaking patterns are usually chosen in
such a way that the symmetric limit reproduces the gross features of
the fermion spectrum (for example only the top quark or the third
family of charged fermions acquire mass) while corrections proportional
to the symmetry-breaking vevs account for a 
parametric suppression of numerically
small  quantities (e.g.~light fermion masses and inter-generational
quark mixings). In the charged fermion sector this strategy is well
justified by the presence of a certain number of $O(1)$ parameters
that are naturally assumed to be non-vanishing in the symmetric
limit. However, because of the exceedingly small value of neutrino
masses, it is conceivable that the whole structure of the neutrino
sector might depend solely on symmetry-breaking effects, in which case
it would be hard to identify a flavour symmetry, even when accurate determinations of 
all the relevant parameters (neutrino masses, mixing, and CP
violating phases)  will be available.\footnote{The
  opposite situation in which  the symmetric limit
  provides a good approximation to the structure of the neutrino mass matrix
  has been recently thoroughly analysed in~\cite{Reyimuaji:2018xvs}.}

As is well known, an elegant explanation for the tininess of 
neutrino masses relies on the assumption that they arise
from  a non-renormalizable effective operator of dimension
five, suppressed by some large scale. Then, if a flavour symmetry
exists, but it is broken above the scale at which the effective
operator is generated, it can remain covert in the structure
of the effective mass operator.  
As we will show, it is however
possible that far-reaching consequences of the symmetry  survive
in the effective theory in the form of non-trivial (although possibly
complicated) relations between low-energy observables.  In this paper
we study this possibility in some detail.  We assume that the neutrino
mass operator is generated by the type-I seesaw mechanism with three 
right-handed (RH) neutrinos, and that the form of the high-energy renormalizable Lagrangian
is determined by a simple $U(1)$ flavour symmetry with
generation-dependent charges.
We assume that the symmetry is spontaneously broken by the vev
$f_\phi$ of a SM singlet field $\phi$, which in our construction is
the only new field besides one additional Higgs doublet and the
three RH neutrinos.  The latter acquire their masses from this
breaking, hence $f_\phi$ is the only high-energy scale.

Much in the spirit of the study carried out in
Ref.~\cite{Bjorkeroth:2018ipq} for the quark sector, we search for
$U(1)$ charge assignments that can reduce the number of independent
parameters to the minimum number compatible with non-vanishing lepton
masses, neutrino mixings, and CP violation, and we argue that such
constructions necessarily involve at least two Higgs doublets.  In the
quark sector it was found that this number matches exactly the number
of SM observables~\cite{Bjorkeroth:2018ipq}, resulting in a complete
determination of the model parameters, but no predictions for the SM
observables. For the lepton sector we find instead a different
result. In order to match our phenomenological requirements, the
high-energy Yukawa sector must involve at least nine real plus one
complex parameter.  However, after integrating out the RH neutrinos,
the number of real parameters needed to describe the charged lepton
(CL) masses and the effective neutrino mass matrix is reduced to six.
Thus, in our construction, the twelve physical observables of the
lepton sector (three CL and three neutrino masses, three mixing angles
and three CP-violating phases) turn out to be related by four
non-trivial conditions that are the low-energy consequences of the
original flavour symmetry.

The paper is structured as follows. In Section~\ref{sec:generalities} we
introduce the basic theoretical framework. In
Section~\ref{sec:minimalmodels} 
we classify the  $U(1)$ flavour  models in terms of different types of mass matrices which are
`minimal' in the sense described above  (the three high-energy matrices  have all together only    
ten non-vanishing entries) and we derive    the relations
between low-energy observables that are the consequence of  the flavour
symmetry.
In Section~\ref{sec:predictions} 
we  confront various textures with the experimental data, identify 
the viable possibilities, and assess their predictions.
In Section~\ref{sec:conclusion} we summarise our findings and draw the conclusions. 
Proofs and ancillary results are given in two Appendices.


\section{General considerations} 
\label{sec:generalities}

A Majorana mass matrix for the light neutrinos can be
described by a $D=5$ effective Weinberg operator, generated at some
large scale $\Lambda_{L} \gg \Lambda_{\rm EW}$ (with $\Lambda_{\rm EW}$ the
electroweak breaking scale) that is the scale suppressing neutrino
masses and lepton number ($L$) non-conserving processes.  We
assume that the renormalizable high-energy couplings of the
uncoloured matter fields respect a generation-dependent $U(1)$ symmetry  
which
is spontaneously broken by a large vev
$f_\phi \gg \Lambda_{\rm EW}$ of a SM singlet scalar field $\phi$.\footnote{This symmetry could be either local or global.
In order to avoid complications with gauge anomalies, we assume for simplicity a global $U(1)$.}
  We have two possibilities:
\begin{itemize}[nosep]
\item[1)] $ f_\phi < \Lambda_L$: the $U(1)$ symmetry is still a good
  symmetry in the effective theory after $L$-breaking. $D=5$
  operators which respect $\ugd$ are initially generated. After $U(1)$
  (and EW) symmetry breaking the structure of the mass matrix will
  still reflect the underlying symmetry in some approximation, e.g.
  entries that are forbidden in the symmetric limit will remain
  suppressed.
  
\item[2)] $ f_\phi > \Lambda_L$: the $U(1)$ symmetry is broken
  already at the level of the renormalizable Lagrangian.  Since there
  is no symmetric limit for the low-energy effective operators,
  their structure cannot directly hint to the $U(1)$
  symmetry. Still, the mass matrices can inherit from the initial
  symmetry a set of nontrivial relations among their  entries.
\end{itemize}
In this paper we explore the second possibility.  For definiteness we
work in the framework of type-I seesaw with three RH neutrinos,
with $\Lambda_L$ corresponding to the RH neutrino mass
scale.%
\footnote{%
  Type-II seesaw models introduce an $SU(2)$ triplet
  $\Delta$ together with a Yukawa term $\bar L^c \Delta L$.  The
  strategy of parameter reduction that we follow in our construction
  is not viable in the the minimal model with a single triplet, but
  might be viable with two or more scalar triplets.  In the type-III
  seesaw the singlet RH neutrinos are replaced by $SU(2)$ triplet
  fermions. In this case our strategy can be straightforwardly
  implemented along the same lines discussed here,  and would yield
  similar results.  It remains to be seen if analogous constructions
  can be implemented in models where the Weinberg operator is
  generated via loop diagrams~\cite{Ma:1998dn}.
}
We assume that  $f_\phi$  is the only  high-energy scale, and that 
the RH neutrino masses are generated by the $U(1)$-breaking vev as
$\Lambda_L = \lambda f_\phi$, with $\lambda$ a generic Yukawa coupling.
For $\lambda \lesssim 1$ the condition
$\Lambda_L \lesssim f_\phi$ is then naturally realised. 
As a guiding
principle, we require that the $U(1)$ symmetry enforces a maximal
reduction in the number of free parameters compatible with the qualitative 
requirements of non-vanishing CL and
neutrino masses, non-vanishing neutrino mixings, and CP violation.
While a vanishing mass for the lightest neutrino is still not ruled out, and CP 
conservation in the lepton sector is still allowed by present
data~\cite{Esteban:2018azc,deSalas:2017kay}, we think that neither of these possibilities 
has strong theoretical motivation and, accordingly, we will not discuss these scenarios.


We denote the RH CL as $e_{\alpha}$, the RH heavy neutrinos as
$N_\alpha$, and the left-handed (LH) lepton doublets as
$\ell_\alpha= (\tilde\nu,e)_\alpha^T$, with $\alpha$
a generation index.\footnote{\label{foot:basis}A tilde over the neutrino fields $(\tilde{\nu})$ 
refers to the neutrinos in the basis in which  the CL  carry 
well-defined $U(1)$ charges, $\nu_\alpha$  ($\alpha = e,\mu,\tau$)  
denotes the neutrino combinations that 
couple to the CL mass eigenstates and defines the `neutrino flavor
 basis',  $\nu'_i$ ($i=1,2,3$) refers to neutrino mass eigenstates.}
To define a flavour symmetry that can act
in the most general possible way, regardless of the type
of field ($N$, $\ell$ or $e$) and of the generation index, we need to
introduce at least two Higgs doublets $H_{1,2}$ with the same weak 
hypercharge $Y=+1/2$ but  with different
$U(1)$ charges $\mX(H_1)=\mX_1$ and $\mX(H_2)=\mX_2\neq \mX_1$.
Since all $U(1)$ charges can be freely redefined by a shift proportional 
to the hypercharge of the corresponding field, we can set $\mX_{1} = - \mX_{2}$. 
To explicitly break the $U(1)^3$ rephasing symmetry of  the kinetic terms of the three scalars 
down to $U(1)^2$ corresponding to hypercharge $\times$ 
flavour $U(1)$, a non-Hermitian coupling  
between $H_{1,2}$ and $\phi$ is needed. There are two inequivalent possibilities corresponding 
to the two renormalizable operators:
\begin{equation}
\label{eq:h12phi}
 H_1^\dagger H_2 \phi, 
\qquad 
 H_1^\dagger H_2 \phi^2. 
\end{equation}
We normalize the charges by choosing $\mX_\phi=2$ so that we can 
label  the two cases by the two different  values of the Higgs charges, namely:
{\bf Case 1}:   $|\mX_{1,2}|=1  $  and {\bf Case 2}:  $|\mX_{1,2}|=2  $.
To avoid  bare EW-invariant mass terms 
we assume that there are no  $U(1)$-invariant RH neutrino bilinears, 
i.e.~$\mX(N_\alpha N_\beta)\neq 0$.
The renormalizable seesaw Lagrangian can then be written as
\begin{equation}
	\mathcal{L} = 
		-  \overbar{e} \lambda^{\!e} \ell\, H_n
		- \overbar{N} \lambda^{\!\nu} \ell\, \tilde{H}_n
		- \frac{1}{2} \overbar{N^c} \lambda^{\!N} N\, \phi^{(*)} 
		+ \mathrm{h.c.} ,
\label{eq:Lren}
\end{equation}
where $\lambda^{\!e}$ and $\lambda^{\!\nu}$ are generic non-singular
Yukawa matrices of complex couplings, $\lambda^{\!N} $ is non-singular
complex symmetric, $H_n= H_1$ or $H_2$ as determined by $U(1)$
invariance, and similarly $\phi^{(*)} = \phi$ or $\phi^*$.  Note that
depending on charge assignments, for each one of the three terms in
\eqnref{eq:Lren}, some couplings can be forbidden, and will correspond to
zero entries in the Yukawa matrices.
Let us introduce the mass matrices 
\begin{equation}
  m_e = \lambda^{\!e} v_n,\quad 
  m_D= \lambda^{\!\nu} v_n,  \quad 
  M_N = \lambda^{\!N} f_\phi , \quad 
  \tilde{m}_\nu = - m_D^T M_N^{-1} m_D \,,
\label{eq:massmatrices} 
\end{equation}
where $v_n=v_1$ or $v_2$ with $v_{1,2} = \braket{H_{1,2}}$  the EW
breaking vevs, $f_\phi=\langle\phi\rangle $, and $\tilde{m}_\nu$ is the mass matrix for the light neutrinos
generated by the seesaw mechanism.  
The matrices  $m_e$ and $\tilde{m}_\nu$ can be written 
in terms of the singular values $\hat{m}_e = \diag(m_e, m_\mu, m_\tau)$ and
$\hat{m}_\nu = \diag(m_1, m_2, m_3)$ as:%
\footnote{%
  We will commonly denote these singular values as `mass eigenvalues'.
}

\begin{equation}
\begin{aligned}
  m_e   &= R \, \hat{m}_e L^\dagger , \\
  \tilde{m}_\nu &= (V \Phi)^* \,\hat{m}_\nu\, (V \Phi)^\dagger ,
\end{aligned}
\label{eq:singular}
\end{equation}
where $R$, $L$ and $V$ are $SU(3)$ matrices and $\Phi$ is a diagonal matrix
of phases chosen in such a way that the singular values of $\tilde m_\nu$ are
real and positive. The mass eigenstate basis for the LH charged
leptons and for the neutrinos is defined respectively as 
$e^\prime_L = L^\dagger e_L $ and
$\nu^\prime = (V \Phi)^\dagger \tilde \nu $, 
and the charged current interaction then reads
\begin{equation}
	\mathcal{L}_{CC} 
	= \frac{g}{\sqrt{2}} \overbar{e_L} \gamma^\mu \tilde \nu W^-_\mu + \text{h.c.}
	= \frac{g}{\sqrt{2}} \overbar{e_L}^\prime \gamma^\mu\, U\, \nu^\prime W^-_\mu + \text{h.c.} , 
\label{eq:Lcc}
\end{equation}
where 
\begin{equation}
	U = L^\dagger V \Phi,
\end{equation}
is the lepton mixing matrix that defines the particular combination
$ \nu_\alpha = U_{\alpha i}\, \nu^\prime_i$ of the three neutrino
mass eigenstates $ \nu^\prime_i$ that couples to a specific CL
$(e^\prime_L)_\alpha$ of flavour $\alpha = e,\mu,\tau$.
The basis $(\nu_e, \nu_\mu, \nu_\tau)$ is generally
referred as the `neutrino flavour basis', and here it represents a third basis
for the neutrino fields. (The basis of mass
eigenstates and the basis in which  the CL fields carry 
well-defined $U(1)$ charges have been introduced in footnote~\ref{foot:basis}.)
The effective neutrino mass matrix 
written in this basis  reads
\begin{equation}
	m_\nu  = L^T \,\tilde{m}_\nu\, L = U^* \,\hat{m}_\nu \,U^\dagger.
\label{eq:flavourbasis}
\end{equation}
We will parametrise the neutrino masses in terms of an overall  scale
$\mu_0$ corresponding to the mass of the lightest neutrino, and of the two
mass-squared differences $\Delta_\odot$ (solar) and $\Delta_\oplus$ (atmospheric)
which have been measured with fair precision in oscillation experiments. For $\mu_0$ only upper 
bounds exist, the most stringent of which comes from cosmological considerations.
This bound is expressed as an upper limit on the the sum of the three 
neutrino masses, and reads $\sum_i m_i\lesssim 0.12-0.60$\,eV, 
depending on which set of data is used~\cite{Aghanim:2018eyx}, however, 
for the purpose of ruling out some types of neutrino mass matrix, 
it will be sufficient for us to impose the much looser bound $\mu_0 \lesssim 1\,$eV.
The mass ordering is also not fully determined: we know the sign of the
solar mass square difference, and we know that
$|\Delta_\odot| \ll |\Delta_\oplus|$, but the sign of the atmospheric
mass difference is yet unknown.  Accordingly, there are two possible
orderings for neutrino masses: $\mu_0 = m_1 < m_2 < m_3$ which is
denoted as normal ordering (NO), and $\mu_0 = m_3 < m_1 < m_2$ which
is denoted as inverted ordering (IO).%
\footnote{Recent fits to the neutrino 
data~\cite{Esteban:2018azc,Capozzi:2017ipn,deSalas:2017kay,Simpson:2017qvj,deSalas:2018bym} 
indicate a preference for NO at about $3\sigma$.} The solar mass squared
difference is conventionally defined as
$\Delta_\odot = m_2^2 - m_1^2 >0 $ and, by defining 
$\Delta_\oplus$ to be also positive, we  have 
$\Delta_\oplus = m_3^2 - m_1^2$ for NO  and 
$\Delta_\oplus = m_2^2 - m_3^2$ for IO, that is:
\begin{equation}
\label{eq:NOIO}
	\hat{m}^2_\nu \equiv\diag(m_1^2,m_2^2,m_3^2) = \left\{ \begin{array}{ll}
		\diag \left(\mu_0^2,\, \mu_0^2 + \Delta_\odot,\, \mu_0^2 + \Delta_\oplus \right)  &\mathrm{(NO)}, \\[4pt]
		\diag \left(\mu_0^2 + \Delta_\oplus - \Delta_\odot,\, \mu_0^2 + \Delta_\oplus,\, \mu_0^2 \right)  ~~&\mathrm{(IO)} .
	\end{array} \right. 
\end{equation}

\section{Minimal seesaw models}
\label{sec:minimalmodels}

Generic seesaw models for lepton masses and mixings usually introduce
a number of free parameters much larger than the number of low-energy
observables.  In this respect, models that only involve the minimum
number of  parameters needed to ensure compatibility with
the observed phenomenology are quite economical, and can thus be considered 
particularly attractive.  Following the same strategy adopted in
Ref.~\cite{Bjorkeroth:2018ipq} to search for minimal models for the
quark sector, we  assume that parameter reduction in the seesaw
Yukawa Lagrangian (\eqnref{eq:Lren}) is enforced by a global $U(1)$ symmetry that
acts on the fermionic fields $\ell, e, N$ and on the scalars 
$H_{1,2}$ and $\phi$, and that forbids the maximum number of entries in
the Yukawa matrices
$ \lambda^{\!e}, \lambda^{\!\nu}, \lambda^{\!N}$ compatibly with
the following requirements:
\begin{enumerate}
\item  There  are no $U(1)$-invariant bilinears of the RH neutrino fields, so that  
$M_N$ depends on a single  high-energy scale $f_\phi$.
  This  implies $\mX(N_\alpha)\neq 0$ and $\mX(N_\alpha)\neq - \mX(N_\beta)$
  for all $\alpha,\beta=1,2,3$.
\item All fermions, including the three light neutrinos, are massive,
  hence $ \lambda^{\!e}, \lambda^{\!\nu}, \lambda^{\!N}$ and the
  corresponding mass matrices $m_e,\,m_D,\,M_N$ in
  \eqnref{eq:massmatrices} have nonzero determinant. This also ensures
  $\det m_\nu \neq 0$.
\item The leptonic Jarlskog invariant
  $J = \impart{U_{11} U_{13}^* U_{31}^* U_{33}}$ is non-vanishing.
  This is equivalent to require that all the neutrino mixing angles
  are non-zero and that the Dirac CP violating phase is non
  vanishing, and implies that at least one complex Yukawa
  coupling cannot be made real by field redefinitions.
\item The way the zero and non-zero entries are arranged in the mass
  matrices must be consistent with some $U(1)$ symmetry, i.e.~it must be possible to 
  enforce the corresponding structures by assigning to the fields  a complete set of $U(1)$ charges.
\end{enumerate}

It is straightforward to obtain the minimum number of non-vanishing
entries in the three Yukawa matrices in \eqnref{eq:Lren} (or equivalently
in $m_e, m_D, M_N$ in \eqnref{eq:massmatrices}) that are required to
satisfy the second and third requirements.  To ensure non-vanishing
determinants we need at least three non-vanishing entries in both $m_e$ 
and $m_D$, which can be taken to lie on the diagonal by a suitable
labeling of the fields. For $M_N$, which is symmetric, two parameters
(three non-zero entries) suffice when they are taken for example  as in
$M_N^{(2)}$ in \eqn{eq:MNx}. 
In this case we need two off-diagonal entries in $m_e$ and/or $m_D$ to generate all the mixings, for 
a total of ten parameters. 
Alternatively, $M_N$ may involve three parameters (five non-zero entries) 
as in the  matrix  $M_N^{(3)}$ in \eqnref{eq:MNx},
\begin{equation}
\label{eq:MNx}
M_N^{(2)}\sim	 \textrm{\MNxx},
\qquad\qquad 
M_N^{(3)}\sim	  \textrm{\MNxxx}.
\end{equation}
In the latter case one off-diagonal entry in $m_e$ or in $m_D$
suffices to ensure non-vanishing mixings, again for a total of ten
parameters. 
It is easy to verify that in both these cases field redefinitions allow us to remove all but 
one complex phase, so that $J \neq 0$. We thus conclude that the 
minimum number of fundamental parameters required to construct high-energy type-I 
seesaw matrices qualitatively compatible with the SM lepton sector (in the 
sense that conditions 2 and 3 are satisfied) is ten.
It is straightforward to verify that, not to exceed this number,  
the following  condition must be  also satisfied:
 \begin{enumerate}
\item[5.] No pair of $N$'s and no pair of $\ell$'s are allowed to have the same charge.
\end{enumerate}
This is  because otherwise $2\times 2$ blocks in a pair of matrices 
(respectively $M_N$ and $m_D$ or $m_D$ and $m_e$) will unavoidably fill up,  
implying a total of at least 12 parameters. 
In particular, this excludes the possibility of having more than three parameters in $M_N$.
 In contrast, a pair of RH leptons 
with equal charges will produce a non-vanishing $2\times 2$ block only in $m_e$, so that 
parameter counting by itself  does not exclude this possibility. 
Condition 5 then explains why at least two Higgs doublets are needed
  to implement schemes with a minimal number of parameters.  Non
  vanishing lepton mixings require that some bilinears involving two
  different $\ell_\alpha$ in the first and/or second term in   \eqn{eq:Lren}   
  couple to a scalar doublet which develops a non vanishing vev.  With
  a single Higgs, this would imply charge relations like $\mX(\bar e_\alpha
  \ell_\beta) = \mX(\bar e_\alpha \ell_\gamma) $ with $\beta\neq
  \gamma$, and hence  $\mX(\ell_\beta)=\mX(\ell_\gamma)$.

In the following we will derive a complete classification of the neutrino mass matrix textures 
that are consistent with the requirement of minimality, 
and comply with the conditions 1-5 listed above. 
This program is carried out in three steps: first we discuss the possible ways to assign the ten 
fundamental parameters to the  matrices $M_N$, $m_D$ and $m_e$.
As a second step,  for each possible assignment  we  study which are the viable {\it structures} for 
the neutrino mass matrix. We use the term {\it structure} to denote the equivalence class 
containing  all those {\it textures} that can be transformed one into  another by permutations 
of the row and column indices. As the third and final step we count how many inequivalent  {\it textures} 
correspond to each structure.  As will become clear below,  here the term  {\it texture}
will have a more general meaning than in its common use,  since besides referring to the 
possible presence of vanishing elements, it will also refer to the presence of specific relations 
among  different entries. 

A clarification  is in order regarding the notation that will be used below:
neutrino mass matrix structures (and textures) only represent 
specific relations that must be satisfied by their elements,  without any 
reference to  numerical values for the non-vanishing entries, which remain unspecified. 
These structures/textures will be denoted by the symbol $\tilde{m}$  for the basis in which 
the CL $U(1)$ charges are well-defined and $m_e$ in general is non-diagonal, 
and  by  the symbol $m$ for the basis in which the CL mass matrix is diagonal. 
This notation keeps the same meaning  as for the matrices  $\tilde{m}_\nu$ and $m_\nu$ 
already introduced  (see \eqn{eq:flavourbasis}), although the subscript ``$\nu$'' is 
dropped. 
The subscripted form, $m_\nu$, will instead refer to the matrix containing experimental numbers.
More precisely, $m_\nu = U^* \hat{m}_\nu U^\dagger$ is obtained by multiplying the  experimentally determined 
PMNS mixing matrix by the diagonal matrix of the neutrino mass eigenvalues $\hat{m}_\nu$
which, for the two cases of normal or inverted ordering (see \eqn{eq:NOIO}) encodes the available 
experimental information on neutrino masses. 
In short,  there are several matrices $m$, one for each texture,  that have to be confronted with 
$m_\nu$(NO) or  $m_\nu$(IO)  to asses whether they can be phenomenologically viable.

\subsection{Neutrino mass structures}
\label{sec:structures}

\begin{table}[t!]
\centering
\begin{tabular}{ccccc}
\toprule
  $M_N$ & $m_D$ & $m_e$ &  &  \\
\midrule
  2 & 5 & 3 & \xmark &  $U(1)$  \\
  2 & 3 & 5 & \xmark & $\theta_{ij}\neq 0$\\
  2 & 4 & 4 & \cmark &   $\;|\mX_{1,2}|= 1,2$ \\  
  3 & 4 & 3 & \cmark &  $|\mX_{1,2}|= 2\;\,   $ \\
  3 & 3 & 4 & \cmark &   $\;|\mX_{1,2}|= 1,2$ \\ 
\bottomrule   
\end{tabular}
\caption{Possible ways to assign the ten parameters of a minimal 
  type-I seesaw model to the mass  matrices. The fourth column
  indicates which parameter assignments can be consistent with a $U(1)$ symmetry 
  and with three non-vanishing mixings. The last column indicates 
  for the non viable cases which condition is violated, and for the remaining cases 
  for which values of the Higgs 
  charges they are viable. 
  }
\label{tab:distributions} 
\end{table}

The possible ways of distributing the ten parameters among $M_N$, $m_D$ and $m_e$
while ensuring non-vanishing determinants for all three matrices,  
are listed in Table~\ref{tab:distributions}. 
The first possibility in the first line is excluded because 
there is no  way of assigning  $U(1)$ charges  to $N_\alpha$ and $\ell_\alpha$  consistently 
with the structure of $M_N^{(2)}$ in such a way that $m_D$ has five non-vanishing entries.
The second possibility can be implemented  consistently with the $U(1)$ symmetry but necessarily
yields some vanishing mixing. This is because  the form of $M_N^{(2)}$ \eqn{eq:MNx}
implies that $m$ has two degenerate eigenvalues, and thus there are unphysical mixings. 
The other three possibilities are viable, although the one on the fourth line is consistent with $U(1)$ 
only in case $|\mX_{1,2}|=2$. 
For each viable configuration listed in Table~\ref{tab:distributions} we now explore the specific form of the viable textures.  
To denote the different possibilities it is convenient to introduce the following compact notation:
each case will be labeled with a string of the form $\mX (n) [i\, j] [k\, l]$.
$\mX=|\mX_{1,2}|=1,2$ distinguishes the two cases corresponding to the scalar couplings in \eqn{eq:h12phi};
$n=2,3$ denotes respectively the structure of the RH neutrino matrix $M_N^{(2)}$ or $M_N^{(3)}$ in \eqn{eq:MNx};
the indices $i,j=1,2,3$ ($i\neq j$) within the first pair of square brackets denote which off-diagonal entry 
in $m_D$ is different from zero  ($m_D^{i\neq j}\neq0 $); 
$k,l$ in the second pair of brackets  have the same meaning but for the charged lepton mass matrix ($m_e^{k\neq l}\neq0 $).
Furthermore, a pair of empty square brackets $[\ ]$  refers to a diagonal matrix,  while 
indices within curly brackets refer collectively to both orderings $\{i\,j\} = [i\,j]$ or $[j\,i]$.
For example:
\begin{equation}
\label{eq:examples}
1(2)[12] \{12\}\!:\ \ \mX\!=\!1,\ \  M_N\!=\!M_N^{(2)},\  
 \ m_D\!\sim\!\textrm{\pmuptw}, \ \ m_e\!\sim\!  \textrm{\pmuptw} \; \textrm{or}  \;
\textrm{\pmdotw}.
\end{equation}
Deriving all the possible textures for the viable parameter assignments  given 
in Table~\ref{tab:distributions} is a bit lengthy, but straightforward. It requires solving for the conditions 
$|\mX(N_i) + \mX(N_j) |=  \mX_\phi$ = 2, 
$|\mX(\bar N_i) + \mX(\ell_j) |=  \mX$, 
$|\mX(\bar e_i) + \mX(\ell_j) |=  \mX$ 
for the specific number of non-vanishing 
parameters in each mass matrix $M_N,\,m_D,\,m_e$,  while respecting the 
constraint  given above at point 5, and ensuring that the total number of non-vanishing 
entries does not exceed ten.  It is worth noting that for $M_N^{(3)}$ the viable RH neutrino charge 
assignment is unique (modulo an overall sign):  $\mX(N)=(1,-3,5)$, and   
this renders the derivation of $\mX(\ell)$ and $\mX(e)$ 
that yield the viable textures
of type $\mX(3)[ij][kl]$ particularly simple.
Conversely, the structure of $M_N^{(2)}$ only implies for the RH neutrino charges the condition $\mX(N) = (1, \mX(N_2),  -\mX(N_2) \pm 2) $. 
Obtaining all textures then requires solving the system of constraints  
for increasing values of $|\mX(N_2)|$ until  no more solutions are found. 

\begin{table}[ht]
\centering
\begin{tabular}{ccccccc}
\toprule
            &  Label                   & $m_e^\dagger m_e$ & $\tilde m$       & $m$    & Viable & Comments \\
\midrule
$\UVX{A}$  & $\mX(2)[21]\{12\}$   & \OneTwoSym        & \mnuArchipelago& \mnuOneTZLower    & \xmark &  \\[2.5ex]
$\UVX{B}$  & $\mX(2)[12]\{12\}$   & \OneTwoSym        & \mnuRibbon    & \mnuOneTZLower    & \xmark & \\[2.5ex]
 $\UVX{H}$           & $1(2)\{23\}\{23\}$         & \TwoThreeSym      & \mnuFarIsland & \TwoThreeSym      & \xmark & $\theta_{ij}\neq 0$ \\[2.5ex]
$\UVX{G}$  & $2(2)[12][13]$             & \OneThreeSym      & \mnuRibbon    & \mnuFull          & \cmark & \\
\midrule
$\UVX{C}$  & $1(3) [~]\{13\}$          & \OneThreeSym      & \mnuIsland    & \mnuOneTZMiddle   & \xmark & \\[2.5ex]
$\UVX{D}$  & $1(3) [~]\{12\}$          & \OneTwoSym        & \mnuIsland    & \mnuFull          & \cmark & \\[2.5ex]
 $\UVX{E}$  & $2(3) [21][~]$            & \NoneNone         & \mnuIsland    & \mnuIsland        & \cmark & Refs.~\cite{MadanSingh2019, Alcaide:2018vni} \\[2.5ex]
 $\UVX{F}$  & $2(3) [31][~]$            & \NoneNone         & \mnuOneTZMiddle& \mnuOneTZMiddle    & \xmark & \\
\bottomrule     
\end{tabular}
\caption{Possible matrix structures for the viable parameter assignments listed in Table~\ref{tab:distributions}. In the first two lines $\mX=|\mX_{1,2}|$ 
is left unspecified, meaning that the same textures are obtained in both cases  $\mX=1$ and  $\mX=2$.  Structures of type  $\UVX{H}$
 yield some vanishing mixing, and are thus ruled out by the condition $\theta_{ij}\neq 0$  as is indicated in the last column.  Structures  of type  $\UVX{E}$
with two vanishing parameters can yield viable models. This type of matrices  has been already thoroughly analysed in the literature 
\cite{MadanSingh2019,Alcaide:2018vni,Singh:2016qcf,Zhou:2015qua,Kitabayashi:2015jdj,%
Fritzsch:2011qv,Meloni:2012sx,Dev:2006qe,Guo:2002ei,Frampton:2002yf}, and hence  are omitted from our analysis.}
\label{tab:textures}    
\end{table}
The viable structures that result from this procedure are listed  in Table~\ref{tab:textures}.
The first half of the table (structures $\UVX{A}$, $\UVX{B}$, $\UVX{H}$, $\UVX{G}$) correspond to the RH neutrino mass matrix  $M_N^{(2)}$, 
which is symmetric under the exchange  $N_2 \leftrightarrow N_3$. This means that structures obtained 
by exchanging the $(2, 3)$ labels of all the fields are also viable. They have not been listed in the table 
because this  becomes redundant in view of the next step of the analysis, that will precisely deal with 
field relabeling. 
By contrast, there is no such symmetry for $M_N^{(3)}$, and thus no structure has been omitted 
in the second half of the table.
In the third column we give the form of the matrix $m_e^\dagger m_e$ from which  
the structure of the LH  CL mixing matrix  $L$ (see \eqn{eq:flavourbasis}) can be read off.  It is important to remark that 
 $L$ is either trivial (structures $\UVX{E}$, $\UVX{F}$) or it corresponds to a simple 
matrix  containing a $2\times 2$ unitary block.  
In the fourth column $\tilde{m}$ gives the structure of the 
neutrino mass matrix in the basis in which the CL charges are well-defined (non-diagonal $m_e$),  while  the fifth 
column gives the form of  the neutrino mass matrix $m$ in the flavour basis, in which  $m_e$ is diagonal
(since for  $\UVX{E}$ and $\UVX{F}$  the matrix $L$ is trivial, in the last two lines  $\tilde m= m$).  
The column labeled ``Viable'' marks whether the structure is compatible (\cmark) or not (\xmark)  with experimental data.

It is readily seen that case $\UVX{H}$ is not viable since the structures  of   $m_e^\dagger m_e$ and of ${m}$ imply that one lepton flavour does not mix 
with the others and thus there are vanishing mixings. 
Structures of type $\UVX{E}$ have two vanishing parameters in $m$. Neutrino matrices of this type have been thoroughly studied in the literature 
\cite{MadanSingh2019,Alcaide:2018vni,Singh:2016qcf,Zhou:2015qua,Kitabayashi:2015jdj,Fritzsch:2011qv,Meloni:2012sx,Dev:2006qe,Guo:2002ei,Frampton:2002yf}
and it is known that they can yield viable models.
More precisely, it is possible to find suitable textures in the sense of specific positions for the zero entries,  and specific values for the four non-vanishing parameters 
that fit  well the experimental values for the three  mixing angles and two mass squared differences, giving predictions for the remaining parameters.
Quantitative analysis of two-zero neutrino mass textures have been performed recently in Refs.~\cite{MadanSingh2019,Alcaide:2018vni}, 
hence we omit this type of structure from our numerical analysis. 
Let us, however, remark that while published numerical studies
generally assume specific textures or specific relations between
neutrino mass entries~\cite{Barbieri:2003qd} from the start, 
 here we have identified in a top-down approach which $U(1)$ symmetries  can produce such textures or, 
in other words, we have pinpointed the complete UV seesaw models in terms of viable sets of $U(1)$  charge assignments. 
For completeness we list them in Appendix~\ref{sec:viablecharges}.\footnote{For structure $\UVX{E}$ 
 there is a one-to-one correspondence between a specific texture for $m$ and the high-energy seesaw mass matrices, 
that is there is a unique string label   (e.g.~$2(3)[21][\, ]$ for the case in Table~\ref{tab:textures}) 
and a unique set of charges identifying each model. It is thus conceivable that the underlying $U(1)$ symmetry could be identified solely from low-energy data. 
Hence these cases do not properly belong to the class of ``covert'' $U(1)$ symmetries.}
For the remaining structures in Table~\ref{tab:textures}, their viability (or not) has been  assessed by means of  the numerical analysis 
described in the next section.

\subsection{Matrix textures and analytical constraints}
\label{sec:textures}

There are two important points that must be addressed before proceeding with the numerical analysis. 
The first regards the number of free parameters in the low-energy effective theory and the relationships between matrix elements.
The second concerns the number of textures, or equivalently of models, obtained from a given low-energy structure by permutations of rows and columns.

Let us address the first point. The high-energy seesaw model depends on ten parameters, but three RH neutrino masses are integrated 
out, so that the effective theory is defined by  seven parameters only.  In the flavour basis, three parameters correspond to the CL masses, 
so  $m$  must have only four independent parameters.  There are three different ways in which this condition can be realised:  $(i)$ two elements of $m$ vanish, as for structures  of type $\UVX{E}$;  $(ii)$ only one element vanishes, as for $\UVX{A},\UVX{B},\UVX{C},\UVX{F}$, and thus  
one relation $f(m_{ij})=0$ between the non-vanishing entries must hold;  $(iii)$ all elements of $m$ are non-vanishing, 
as for $\UVX{G},\UVX{D}$, and then two relations must hold. The following results are derived by starting from an explicit form  
for $M_N$ and $m_D$ in terms of symbolic entries,  deriving  $\tilde m$ according to the last relation in \eqn{eq:massmatrices}, 
writing down $m = L^T \tilde m L$ where the nontrivial $2\times 2$ block in $L$ is written as a generic $SU(2)$ matrix, 
and  finally inspecting the resulting expression for $m$ to identify the sought relations. 

Structures of type $\UVX{E}$ have four non-vanishing entries that are therefore all independent. 
We can rephrase this by stating that  textures belonging to class $\UVX{E}$ must satisfy the 
two conditions
\begin{equation}
\label{eq:twozerorel}
(\UVX{E}):\  \left\{
  \begin{aligned}
  m_{\bar k \bar k} &= 0, \\
  m_{\bar k j} &= 0, \quad  (j \neq \bar k) , 
  \end{aligned}
\right.
\end{equation}
where the barred index $\bar k$ must match between the two equations. 
This index-matching implies that, within our minimal scenario defined by $U(1)$  
charge consistency and two leptonic Higgs doublets, the two texture zeros 
must appear in the $2\times 2$  block with rows and columns $j$ and $\bar k$. This excludes for example putting the zeros in 
the anti-diagonal, and analogous textures obtained by index permutations, which instead 
have been regularly considered in the literature on two-texture zero matrices. 

In the flavour basis, structures of type $\UVX{A}$ and  $\UVX{B}$ are the same,  and  $\UVX{F}$ is equal to  $\UVX{C}$. 
Moreover the latter two can be obtained from the former by permuting the $(2, 3)$ indices.   
These structures have a single (diagonal) zero entry, and thus, as anticipated, there must be one non-trivial relation between 
the other non-vanishing entries. Analysing the expression of $m$ in terms of explicit  
seesaw parameters this relation is easily derived, and it corresponds to the second  of the following two conditions: 
\begin{equation}
\label{eq:onezerorel}
(\UVX{A}, \UVX{B}, \UVX{C}, \UVX{F}):\  \left\{
  \begin{aligned}
    m_{\bar k \bar k} &= 0,  \\
    m_{ii} m_{jj} - m_{ij}^2 &= 0 , \quad  (i,j\neq \bar k),
  \end{aligned}
  \right.
\end{equation}
where $\bar k$ must be chosen to match the unmixed entry in the CL mass matrix.
Finally, structures $\UVX{G}$ and $\UVX{D}$, which are related by permutation of the $(2,3)$ indices (see Table~\ref{tab:textures}), have no zero entries, 
and thus there must be two non trivial relations between their six non-vanishing parameters. These two conditions are: 
\begin{equation}
\label{eq:nozerorel}
(\UVX{G},\UVX{D}):\  \left\{
  \begin{aligned}
    m_{ii} m_{jj} - m_{ij}^2 &= 0, \\ 
    \sum_{\ell = i,j} \left( m^*_{\ell \bar k} \sqrt{m_{\ell\ell} m_{\bar k\bar k}} 
      - |m_{\ell\ell} m_{\bar k\bar k}| \right) &=0, \quad  (i,j\neq \bar k),
  \end{aligned}
\right.
\end{equation}
where again $\bar k$ must match the index of the unmixed entry in the CL mass matrix ($\bar k = 2$ for $\UVX{G}$ and $\bar  k = 3$ for $\UVX{D}$).

Let us now briefly comment on the number of independent CP-violating rephasing invariants.
A generic  $3\times3$ Majorana neutrino mass matrix has three independent 
invariants~\cite{Nieves:1987pp} that are usually expressed in terms of elements of the PMNS mixing matrix, 
but that can be also expressed in terms of elements of the effective neutrino mass matrix $m$~\cite{Sarkar:2006uf}.
We focus here on case ($\UVX{G}, \UVX{D}$) since it is this structure that eventually will yield the most interesting model. 
In the absence of zero textures, the expression for the rephasing invariants has the    
particularly simple form 
$J_{ij} = J_{ji} = \mathrm{Im} (m_{ii} m_{jj} m_{ij}^*  m_{ji}^*)$  
\cite{Sarkar:2006uf}.
Taking $i$ and $j$ as defined in \eqn{eq:nozerorel} and writing 
$J_{ij} = \mathrm{Im}(m_{ii}m_{jj})\,\mathrm{Re}(m_{ij}^2) - \mathrm{Re}(m_{ii}m_{jj})\,\mathrm{Im}(m_{ij}^2) $, 
it is easy to see that 
the first condition in \eqn{eq:nozerorel} implies  $J_{ij}=0$.  
Taking instead the indices $i$ and $\bar k$, and writing 
$J_{i\bar k} = 2\, \mathrm{Im} (m_{i\bar k}^* \sqrt{m_{ii}m_{\bar k \bar k}})\, \mathrm{Re} (m_{i\bar k}^* \sqrt{m_{ii}m_{\bar k \bar k}}) $,
the second condition in \eqn{eq:nozerorel} implies 
$\sum_{i\neq \bar k}  \mathrm{Im} (m_{i\bar k}^* \sqrt{m_{ii}m_{\bar k \bar k}})=0 $.
From this one can derive the relation 
$ J_{j\bar k}  = - \frac{\mathrm{Re} (m_{i\bar k}^* \sqrt{m_{ii}m_{\bar k \bar k}}) }{\mathrm{Re} (m_{j\bar k}^* \sqrt{m_{jj}m_{\bar k \bar k}}) } J_{i\bar k}$.
Therefore there is only one independent, non-zero CP-violating rephasing invariant. This result can be traced back 
to the fact that with ten fundamental seesaw parameters there is only a single physical CP violating phase, and implies 
that although  in general   for these structures the  Dirac phase $\delta$ as well as the two Majorana phases 
will be non-vanishing, they must all be related.

The second point that we have to address  regards the number of different textures that correspond to each 
structure listed in Table~\ref{tab:textures}. 
It is clear that to be able to confront the various textures  
with experimental neutrino data, we have to decide an ordering for the lepton fields, to give a precise
meaning to the flavour labels $e, \mu, \tau$.
Most convenient of course is to fix the ordering in $\hat m_e$ from light to heavy, so that the entries 
in the neutrino mixing matrix will keep their usual meaning, and count how many inequivalent 
textures can be obtained by permuting the rows and columns in $m$. Since $m$ is symmetric, there 
are in principle six possibilities corresponding to permutations of the indices $1,2,3$. 
With reference to \eqn{eq:twozerorel}, for structure $\UVX{E}$ we have three ways to choose the index $\bar k$ corresponding to the texture zero on the diagonal in $m$ (the first condition). For each of these, there are two ways to chose the index $j$ in the 
second condition, for a total of six inequivalent textures.  
For structures ($\UVX{A}, \UVX{B}, \UVX{C}, \UVX{F}$) there are again three ways to choose the index $\bar k$ of the vanishing entry in $m$.
However, since the second condition in \eqn{eq:onezerorel} is symmetric under the exchange of 
the other two indices, $i\leftrightarrow j$, these permutations are equivalent.
It is thus sufficient to fix $\bar k$ to identify the three inequivalent textures.
Structures ($\UVX{G}, \UVX{D}$) behave similarly: both conditions in \eqn{eq:nozerorel} 
are symmetric under the exchange $i\leftrightarrow j$, hence once $\bar k$ is fixed, the three textures, which in this case refer to the way the indices $1,2,3$ are assigned to $i,j, \bar k$ in \eqn{eq:nozerorel}, remain univocally identified.

As anticipated, in the numerical study we will omit an analysis of the two-zero textures of structure $\UVX{E}$ since thorough studies already exist in the literature.
The three textures belonging to structures ($\UVX{A}, \UVX{B}, \UVX{C}, \UVX{F}$) have one vanishing entry in $m$, and 
henceforth will be collectively referred to as textures of ``type {\bf 1}'' .
The three textures in structures ($\UVX{G}, \UVX{D}$) have no zero texture, and will be collectively referred as textures  of ``type {\bf 0}''.
For both types {\bf 1} and {\bf 0},  a texture can be identified with one of the following three labels 
\begin{equation}
	[ij\bar k] \in \left\{ [123], ~[132], ~[231] \right\},
\label{eq:ijk}
\end{equation}
where we have chosen for definiteness $i<j$.

\section{Confronting textures with neutrino data}
\label{sec:predictions}

In this section we study whether any of the neutrino mass textures $m$ 
belonging to the structures of type {\bf 1} or {\bf 0}  can be in agreement with the available 
low-energy neutrino data.  These data are encoded in the matrix 
$m_\nu = U^* \hat m_\nu U^\dagger$,  with  $U$ the PMNS matrix 
which includes the available data on neutrino mixing, while $\hat m_\nu$ 
is taken to be in one of the two forms given in \eqn{eq:NOIO},  with $\Delta_\odot$ and $\Delta_\oplus$ 
experimentally known and $\mu_0$ bounded from above. 
More explicitly, we need to verify if, given $U$ and $\hat m_\nu$ (in either NO or IO form),  
any of the six pairs of relations obtained from \eqn{eq:onezerorel} and \eqn{eq:nozerorel} by index 
permutations can be satisfied when replacing $m \to m_\nu$(NO) or $m \to m_\nu$(IO). 

\subsection{Analytical expressions for the Majorana phases}
 \label{sec:majoranaphases}

The neutrino mixing matrix is defined as: 
\begin{equation}
\begin{aligned}
	U
		&= R_{23} U_{13} R_{12} P_\varphi \\
		&= 
		\pmatr{1 & 0 & 0 \\ 0 & c_{23} & s_{23} \\ 0& -s_{23} & c_{23}}
		\pmatr{c_{13} & 0 & s_{12} e^{-i \delta} \\ 0&1&0 \\ -s_{13} e^{i \delta} & 0 & c_{12}}
		\pmatr{c_{12} & s_{12} & 0 \\ -s_{12} & c_{12} & 0 \\ 0&0&1} 
		\pmatr{e^{-i \varphi_1/2} &0&0 \\ 0& e^{-i \varphi_2/2} &0 \\ 0&0&1} .
\end{aligned}
\label{eq:PMNS}
\end{equation}
where in the second line $c_{ij}= \cos\theta_{ij}$ and $s_{ij}=\sin\theta_{ij}$.  
The two orthogonal matrices $R_{23}$ and $R_{12}$ and the unitary matrix $U_{13}$ 
respect the standard parametrization. It is, however, convenient for us 
to parametrize the Majorana phases as in \eqn{eq:PMNS}, which differs from 
the more common form  $P_\alpha = \diag(1,e^{i \alpha_{21}/2},e^{i\alpha_{31}/2})$
adopted for example by the Particle Data Group~\cite{Tanabashi:2018oca}. 
The two parametrizations are related by 
$\alpha_{21} = \varphi_{1}-\varphi_{2}$ and $\alpha_{31}=\varphi_1$  up to an irrelevant overall phase.
After substituting  $m \to m_\nu$, each pair of constraints for the two types of textures, \eqn{eq:onezerorel} and \eqn{eq:nozerorel},  
imply four relations (two for the real and two for the imaginary parts of the equations) 
between the nine low-energy observables $\theta_{12}$, $\theta_{13}$, $\theta_{23}$, $\delta$, $\varphi_1$, 
$\varphi_2$, $\mu_0$, $\Delta_\odot$ and $\Delta_\oplus$. In principle this would allow us to express 
four parameters in terms of the remaining five, but in practice the conditions are highly non-linear and involve 
periodic functions, and deriving explicit expressions for the functional dependence 
of some parameters from the others is not possible. 

There is, however, an alternative approach that allows us
to solve for the two Majorana phases analytically in terms of $\delta$ and of the mixing angles $\theta_{ij}$.  
We outline the procedure using as an example texture 
$\UVX{D}$ in Table~\ref{tab:textures}, reproduced here for convenience:
\begin{equation}
\label{eq:Dphase}
\UVX{D}: \quad 
  m_e^\dagger m_e \sim \text{\OneTwoSym} ,  \quad  \tilde m \sim  \text{ \mnuIsland} , 
\end{equation}
where, as it can be checked explicitly, the entries in $\tilde m$ satisfy the relation $\tilde m_{11}  \tilde m_{33}= \tilde m_{13}^2$.  
Let us choose a phase convention such that $(m_e^\dagger m_e)_{12} = (m_e^\dagger m_e)_{21}^*$ is the only complex 
entry, so that $\tilde m$ is real symmetric, and can thus  be diagonalised by an orthogonal
matrix $O$. Multiplication by a diagonal matrix of phases $P_\xi$ is also needed to render all the eigenvalues  positive. 
It is shown in Appendix~\ref{sec:signature} that, in the chosen phase convention, for matrices of the form of $\tilde m$ in \eqn{eq:Dphase}
the eigenvalues with the  largest and smallest  absolute values have the same sign, and opposite 
with respect to the sign of the middle eigenvalue. 
By ordering the absolute values  from small to large,  without loss of generality  (see Appendix~\ref{sec:signature}) the signature of the eigenvalues 
can then be taken as $(+,-,+)$. Therefore $P_\xi = \diag(1,i,1)$.  
In our basis, the matrix that diagonalizes the neutrino matrix $m$ can then be written as $U_\xi = L^\dagger O P_\xi$,
with $L$ the CL mixing matrix. 
Let us now focus on the third row of $U_\xi$, which is unaffected by the matrix $L$ since it is non-trivial only in the upper-left $2\times2$ block. 
Since $O$ has only real entries, we can readily see that $U_\xi(3,1)$ and $U_\xi(3,3)$ are real, while $U_\xi(3,2)$ is purely imaginary.
We confront this result with the corresponding third row of the PMNS matrix. Since the phase conventions for  $U$ and $U_\xi$ 
are different, to carry out the comparison we need  to allow for a   relative phase redefinition.
This can be done by multiplying $U$ by a generic  diagonal matrix of 
phases $P_\rho= \diag(e^{i\rho_1}, e^{i\rho_2},e^{i\rho_3})$. Direct comparison between 
$(P_\rho U)_{3j}$  and $(U_\xi)_{3j}$  for the three elements $j=1,2,3$ yields the following conditions: 
\begin{equation}
\begin{aligned}
\label{eq:phases}
  \mathrm{Im}(P_\rho U)_{31} &=  \mathrm{Im}\left[e^{i(\rho_3-\frac{\varphi_1}{2})} (-e^{i\delta} c_{12} c_{23} s_{13})+s_{12} s_{23}\right] = 0 , \\
  \mathrm{Re}(P_\rho U)_{32}  &=   \mathrm{Re}\left[e^{i(\rho_3-\frac{\varphi_2}{2})} (-e^{i\delta} s_{12} c_{23} s_{13})-c_{12} s_{23}\right] = 0 , \\
  \mathrm{Im}(P_\rho U)_{33}  &=   \mathrm{Im}\left[e^{i\rho_3} c_{13}c_{23}\right] =0 .  
\end{aligned}
\end{equation}
The last equation fixes $\rho_3 = 0 $ or $\pi$, and  we can then obtain $\varphi_1$ and $\varphi_2$ from the other two equations
in terms of $\delta$ and $\theta_{ij}$. The same procedure can be carried out for all the other textures.  The final result 
depends on the particular texture $[ij\bar k]$ since different rows of the PMNS matrix $U$ are selected for the comparison 
for the different cases, but it does not depend on whether the texture is of type {\bf 1} or {\bf 0}.  We obtain: 
\begin{equation}
	\tan \left( \frac{\varphi_1}{2} \right) = \frac{\tan \delta}{\Omega^{(\varphi_1)}_{[ij\bar k]}}, \qquad
	\tan \left( \frac{\varphi_2}{2} \right) = - \frac{\Omega^{(\varphi_2)}_{[ij\bar k]}}{\tan \delta} , 
\label{eq:majoranaphasesexact}
\end{equation}
with 
\begin{equation}
\begin{aligned}
	\Omega^{(\varphi_1)}_{[123]} &= 1 - \frac{t_{12} t_{23}}{s_{13} \cos \delta }, & 
	\Omega^{(\varphi_2)}_{[123]} &= 1 + \frac{t_{23}}{t_{12} s_{13} \cos \delta}, \\
	\Omega^{(\varphi_1)}_{[132]} &= 1 + \frac{t_{12}}{s_{13} t_{23}\cos \delta }, \qquad& 
	\Omega^{(\varphi_2)}_{[132]} &= 1 - \frac{1}{ t_{12} s_{13} t_{23}\cos \delta}, \\
	\Omega^{(\varphi_1)}_{[231]} &= 1, & 
	\Omega^{(\varphi_2)}_{[231]} &= 1, \\
\end{aligned}
\label{eq:majoranaphasesOmega}
\end{equation}
where $t_{ij} = \tan\theta_{ij}$ and $s_{ij} = \sin\theta_{ij}$. 
Note that for the texture $[231]$ the phase relations are particularly simple: $\varphi_1 = \varphi_2 - \pi = 2\delta$.

\subsection{Numerical analysis}
\label{sec:minimisation}

The possibility of writing down  analytic relations for the Majorana phases in terms of the
other low-energy parameters is a nice result, but given that $\varphi_{1,2}$ are the 
most difficult quantities to determine experimentally, in practice there is little hope to directly test  
the model via  \eqn{eq:majoranaphasesexact}.   Still, it is useful to eliminate $\varphi_{1,2} $ 
from  \eqn{eq:onezerorel} and \eqn{eq:nozerorel} by expressing them in terms  of the 
other parameters.  
Note, however, that in doing so we implicitly reduce the number of constraints, since 
the existence of matrix textures that yields the relations 
\eqn{eq:onezerorel} and \eqn{eq:nozerorel} is also what has been used to determine  
the functional dependence of $\varphi_{1,2} $ on $ \delta$ and $\theta_{ij}$.
 In short, although the number of 
independent parameters has now dropped from nine to seven, the number of independent 
constraints represented by each pair of complex conditions has also (implicitly) dropped from four to two.
Thus only two additional quantities can  be determined in terms of the others five by solving for the two real 
and two imaginary conditions for each one of the two cases.
As we have already mentioned, it is not possible to analytically extract from 
\eqn{eq:onezerorel} and \eqn{eq:nozerorel} the dependence of any pair of 
parameters from the others, and thus we need to resort to numerical methods. 

We proceed as follows: respectively for textures of type {\bf 1} and {\bf 0}, we combine the pair of conditions in a single expression:
\begin{align}
\label{eq:cond1}
	\cond_\strucA&= \eta^{-4} \left| \mfb_{ii} \mfb_{jj} - \mfb_{ij}^2 \right|^2 + \eta^{-2} \left| \mfb_{\bar k\bar k} \right|^2, \\
	\label{eq:cond0}
	\cond_\strucB &= \eta^{-4} \left|\mfb_{ii} \mfb_{jj} - \mfb_{ij}^2 \right|^2 
+\eta^{-4}\left|\sum\nolimits_{\ell \neq \bar k} \left( \mfb^*_{\ell \bar k} \sqrt{\mfb_{\ell\ell} \mfb_{\bar k\bar k}} 
      - |\mfb_{\ell\ell} \mfb_{\bar k\bar k}| \right)\right|^2,
\end{align} 
where
\begin{equation}
	\eta^2 = \tr(\mfb^\dagger \mfb) = \sum\nolimits_{i}\mu_i^2\,,
\end{equation}
is a normalisation factor that ensure that $\cond_{\strucA,\strucB}$ are dimensionless and, not to clutter the notations,  we have omitted 
the label $\nu$ from the matrix elements: $(m_\nu)_{ij} = m_{ij}$.   
Each matrix element $m_{ij} = m_{ij}\left(\{p\}\right)$ depends on the seven parameters
 $\{p\} = \{\theta_{12}, \theta_{13},\theta_{23},\delta,\mu_0,\Delta_\odot,\Delta_\oplus\}$, and so do the
 two functions $\cond_{\strucA,\strucB}  = \cond_{\strucA,\strucB} \left(\{p\}\right)$. We have to search for values 
 of these parameters that yield $\cond_{\strucA,\strucB}=0$. 
Since $\cond_{\strucA,\strucB}$ depend only on the absolute values of the conditions
\eqns{eq:onezerorel}{eq:nozerorel}, $\cond_{\strucA,\strucB} = 0$ necessarily corresponds to a global minimum,
and thus the search can be carried out via numerical minimization. 
Let us note at this point  that  $\delta \to -\delta$   corresponds to  $U\to U^*$  which in turn implies  $m_{ij} \to m^*_{ij}$.
 However, under  complex conjugation  all the conditions in \eqns{eq:onezerorel}{eq:nozerorel}   remain invariant, 
so that we cannot expect to obtain information about the half-plane in which $\delta$ lies.
In carrying out the minimisation, it is convenient to restrict the parameter space $\{p\}$ by  
excluding regions that are experimentally known to be highly unlikely.
This can be done, for each numerical run, by sampling the values of the most precisely 
measured quantities from normal distributions with mean and variance corresponding to 
their  experimental values. In principle we could consider  $\Delta_\odot$, $\Delta_\oplus$, $\theta_{12}$, $\theta_{13}$ 
and $\theta_{23}$ as measured, and derive from minimization the values of the neutrino mass scale $\mu_0$ 
and of $\delta$.  However,  global fits to neutrino data indicate that the shape of  $\chi^2$  for  $\theta_{23} $ is far from parabolic, 
resembling rather a bimodal distribution, with two different minima respectively located in the first and 
second octant (see e.g.~\cite{Esteban:2018azc}), so that using a normal distribution for this parameter is not justified. 
We have then opted for  minimizing with respect to the three variables $\mu_0,\delta$ and $\theta_{23}$, since in this way we   
can also  put in evidence their reciprocal  functional  dependence.

\subsection{Results}

\def\lcmark{{\Large \cmark }}
\def\lxmark{{\Large \xmark }}
\def\sxmark{{\footnotesize \xmark }}

\def\lsim{\raise0.3ex\hbox{$\;<$\kern-0.75em\raise-1.1ex\hbox{$\sim\;$}}}

\begin{table}[t!]
\centering
\begin{tabular}{ccccccc}
\toprule
    \raise-3pt\hbox{No. of zero } &\multicolumn{3}{c}{Normal Ordering} & \multicolumn{3}{c}{Inverted Ordering} \\
\cmidrule(lr){2-4} \cmidrule(lr){5-7}   
    \raise3pt\hbox{  textures} & $[123]$ & $[132]$ & $[231]$  & $[123]$ & $[132]$ & $[231]$ \\
\midrule
    $\mathbf{1}$ & \sxmark & \sxmark & \lxmark & \sxmark & \sxmark & \lxmark \\
    $\mathbf{0}$ & \lxmark & \lcmark & \lxmark &\lxmark & \lxmark &\lxmark \\
\bottomrule
\end{tabular}
\caption{ 
Summary of the numerical results.  A texture labeled with a large x-mark  (\lxmark) indicates that the  conditions  
    $\cond_{\strucA,\strucB} =0$  cannot be satisfied.    A small x-mark  (\sxmark) indicates that 
     $\cond_{\strucA,\strucB} =0$ can only be obtained for values  of $\theta_{23}$ or $\mu_0$ 
     that are experimentally excluded.  The  checkmark \lcmark for 
     the  texture  $[132]$  with no texture zero highlights  the only viable possibility.  
}
\label{tab:results}
\end{table}

The complete numerical analysis requires  twelve numerical minimizations 
corresponding to  three textures $[ij\bar k]$ for each one of the two structures  $\strucA$ and $\strucB$, that have to be confronted with neutrino data assuming in turn NO or IO. For each case there are three possible outcomes: 
\begin{enumerate}[nosep]
\item
 There is no set of values for $\mu_0,\, \delta$ and $  \theta_{23}$  that can  yield $\cond = 0$. These cases are  
 thus inconsistent with the constraints and can be immediately discarded.  The  seven possibilities  marked with 
 a large x-mark (\lxmark)  in Table~\ref{tab:results} belong to this class.  This includes all models with $[ij\bar k] = [231]$.
\item
 There exists a region where $\cond = 0$, but the required values of the parameters are incompatible with 
 experimental limits. Hence also these possibilities are ruled out.  Four  cases belong to  this class, 
 and are marked in the table with a small x-mark (\sxmark). 
 They correspond to the two texture $[123]$ and $[132]$ for structure $\strucA$. When confronted with NO,  $\cond_\strucA = 0$ requires    $\theta_{23} \sim  \frac{2 \pi}{5}$ for $ [123]$ and 
 $\theta_{23} \sim \frac{\pi}{10}$ for $[132]$.  When confronted with IO, for both textures   $\mu_0 \gg 1$ eV is required 
 to satisfy the condition.
 \item
 If $\cond = 0$ is instead satisfied for values of the parameters within the experimental limits, the texture is viable.  
 We find only one viable case, that is highlighted  in the table with a check-mark (\lcmark).   
 It corresponds to texture $[132]$  for a neutrino mass matrix structure without  texture zeros,  that 
  agrees well with all neutrino data when NO is assumed.  We label this case $[132]_{\mathbf{0}}$ . 
 \end{enumerate}
 
\begin{figure}[t!]
\centering
	\includegraphics[width=\textwidth]{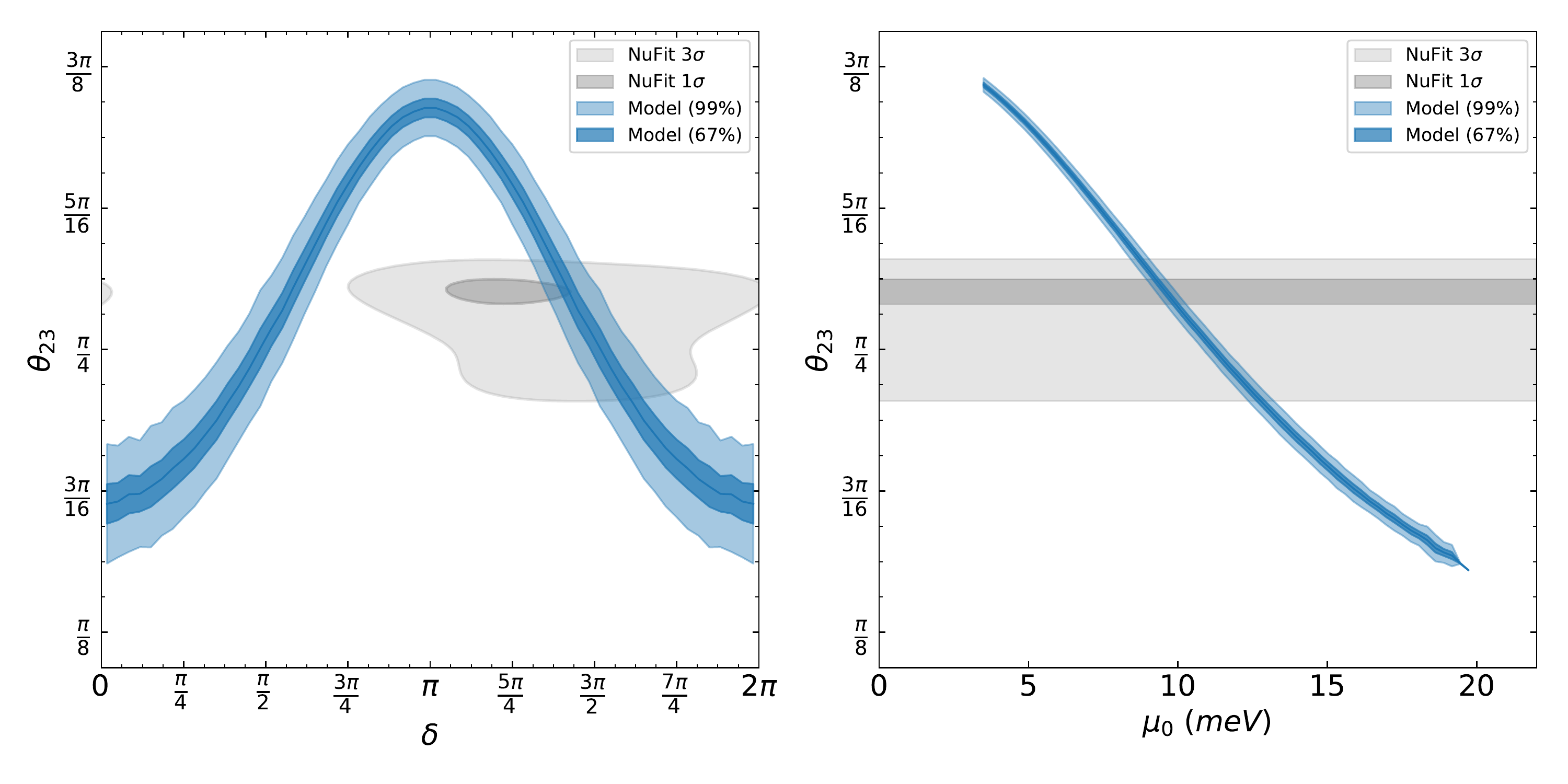}
	\caption{Left panel:  functional dependence of $\theta_{23}$ vs.~$\delta$
at the  $\cond_\strucB = 0$ absolute minimum.~The  dark (light) gray region depicts the 
$1\sigma$ ($3\sigma$) experimentally allowed limits  from Ref.~\cite{Esteban:2018azc}. 
The meaning of the dark and light blue regions is explained in  the  text. 
Right panel: same  for $\theta_{23}$ vs.~$\mu_{0}$. 
}
\label{fig:twoplots}
\end{figure}
 
\begin{figure}[t!]
\centering
  \includegraphics[width=0.7\textwidth]{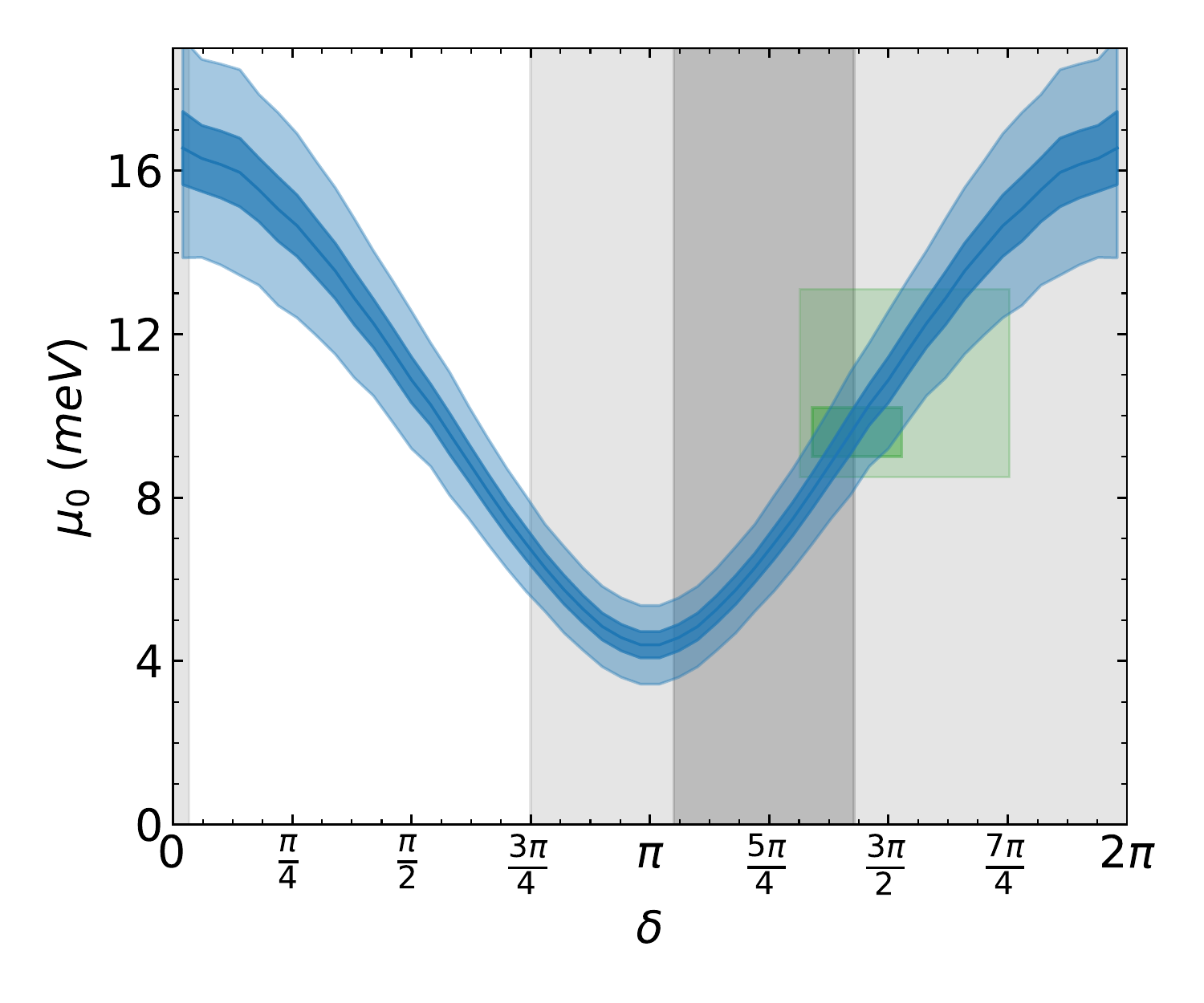}
\caption{
Functional dependence of  $\mu_{0}$ vs.~$\delta$ 
at the  $\cond_\strucB = 0$ absolute minimum. 
The dark (light)  gray area corresponds to the $1\sigma$ ($3\sigma$)
experimentally allowed region for $\delta$  from Ref.~\cite{Esteban:2018azc}. 
The dark (light)  green area  encloses the values of $\delta$ and $\mu_0$ 
for which the corresponding values of $\theta_{23}$  (see  the two plots in Fig.~\ref{fig:twoplots}) 
remain  within the  $1\sigma$  ($3\sigma$ ) experimental limits. 
}
\label{fig:oneplot}
\end{figure}

Besides being viable, texture  $[132]_{\mathbf{0}}$  also yields rather precise predictions, which 
are  depicted  in figs.~\ref{fig:twoplots} and \ref{fig:oneplot}.
Let us explain how these plots are obtained and their meaning.
For  a given set of reference values for  $\Delta_\odot$, $\Delta_\oplus$, $\theta_{12}$, $\theta_{13}$,  the equation 
$\cond_\strucB=0$ represents a line in   the $(\delta,\,\mu_0,\,\theta_{23})$  3D space.  
When the reference values  are allowed to vary within their experimental errors, this line is replaced by 
 a volume.  We  construct this volume by generating  a large set of $\cond_\strucB=0$  points 
 for which the reference values  are drawn from  the experimental distributions for  the corresponding parameters, assumed 
 to be normal and uncorrelated. The points are then binned according to their value of $\delta$, and for each bin  
 the mean $\mu$ and standard deviation $\sigma$ for the values of  $\theta_{23}$ and $\mu_0$  within that bin are 
 computed. We then draw two volumes, corresponding respectively to $\mu \pm \sigma$ (dark blue) and 
 $\mu \pm 3 \sigma$ (light blue). 
 Fig.~\ref{fig:twoplots} depicts the projection of these volumes on the two planes  $(\delta,\theta_{23})$ 
 (left panel) and  $(\mu_0,\theta_{23})$ (right panel), that is the   
marginalisation of  the functional dependences  $\theta_{23}=\theta_{23}(\delta)$ and  $\theta_{23}=\theta_{23}(\mu_0)$
respectively over $\mu_0$ and $\delta$.
Of course  these regions should not be interpreted as representing an experimental statistical significance,
but they still give a qualitatively meaningful  account of the  accuracy with which  the  functional dependences can be determined.
For example,  from the plot on the right  we see that   $\theta_{23}$ and $\mu_0$ are approximately 
linearly anticorrelated, and that the  spread around the central line  is rather small. 
An interesting  prediction  that can be read out from the left plot is that approximate maximal atmospheric mixing
$\theta_{23} \approx \frac{\pi}{4}$   favours  nearly maximal  CP violation
$\delta \approx \frac{\pi}{2}, \frac{3\pi}{2}$.   In both plots the dark and the light grey regions depict respectively 
 the $1\sigma$ and $3\sigma$ experimental  uncertainties on $\theta_{23}$  and $\delta$, taken 
 from Ref.~\cite{Esteban:2018azc}. 

In Fig.~\ref{fig:oneplot} we plot the functional dependence 
between $\mu_{0}$ and $\delta$ at the absolute minimum $\cond_\strucB = 0$.
The gray regions  correspond to the $1\sigma$ (dark) and  ($3\sigma$) (light) 
experimentally allowed limits for $\delta$  from Ref.~\cite{Esteban:2018azc}.
The dark (light) green box encloses the values of $\delta$ and $\mu_0$  for which the 
corresponding values of $\theta_{23}$ in the two plots in Fig.~\ref{fig:twoplots} remain  
within the $1\sigma$  ($3\sigma$) experimental limits.
The vertical sides of the  boxes are  located at 
$\delta = 4.45_{-0.24\, (0.32)}^{+0.35\, (1.06)}$ rad.  
We see that the prediction of the model   for $\delta$ in terms of the experimentally allowed values of $\theta_{23}$
is in good agreement with the direct determination from the global fit in Ref.~\cite{Esteban:2018azc}.
The horizontal sides are located at    $\mu_{0} = 9.5_{-0.5\, (1.0)}^{+0.7\, (3.6)}$ meV,   
and  provide a rather precise prediction for the neutrino mass scale.  
There is in principle a second set of green  boxes that can be obtained by reflection with respect to 
the $\delta=\pi$ vertical line. This is because, as mentioned above, the texture conditions 
do not distinguish between values of $\delta$ in the first or in the second half plane. 
However, values of $\delta$ centred around $\frac{\pi}{2}$ are disfavoured by direct measurements, 
so that we have not drawn the corresponding boxes. 

\section{Concluding remarks}
\label{sec:conclusion}

The quest for a structure of the neutrino mass matrix that can agree with all the available neutrino data 
and at the same time  yield testable predictions,  has represented a major effort pursued in the last two decades 
by many theorists.
Predictivity is obviously linked to a reduction in the number of free parameters, and a simple way to achieve this is to 
assume that some entries in the neutrino mass matrix vanish. For this reason  neutrino mass matrix  structures with one 
or two vanishing entries have been thoroughly studied in the literature. 
In this paper we have adopted a top-down approach.
We have assumed that the neutrino mass matrix is described by an effective Weinberg operator that arises from the type-I seesaw, 
and that a $U(1)$  flavour symmetry exists which determines the structure of the renormalizable high-energy Lagrangian.
As a first step we have searched  for $U(1)$ symmetries able to reduce the number 
of seesaw parameters  to the minimum number required to account qualitatively for three non-zero neutrino masses 
and mixing angles while allowing  for CP violation in the lepton sector. We have given a complete classification of  the various possibilities 
that can arise when the number of leptonic Higgs scalars carrying different $U(1)$ charges is restricted to two. 
 We have then identified the corresponding mass matrix structures, as well as the different textures that can arise in the 
 low-energy effective theory. We have found that in some cases mass matrices with one or two zero textures do arise, 
 corresponding to the condition  $m_{ij}=0$ for some values of the indices. 
However,  more generally these symmetries can produce complicated 
 conditions  involving several entries of the mass  matrix at the same time, 
 that can be written as $f(m_{ij})=0$ with $f$  some well-defined  function.  
 These conditions have the same effect of  zero-textures in reducing the number of parameters, 
 as they allow one to express (although only implicitly)  some parameters in terms of the others. 
 We have then extended the meaning of the word `texture' to refer also to the possibility of exact relations between the parameters. 
 We have found that some cases with two-zero textures can indeed be generated by a $U(1)$  
 symmetry, while others that have also been considered  in the literature cannot, at least in the simplest case of two leptonic Higgs doublets.  
 Only one of the matrices with two-zero textures we have found is viable, 
 as it has been  confirmed by recent  up-to-date numerical analysis~\cite{MadanSingh2019,Alcaide:2018vni}. 
 Textures with a single vanishing entry are also  generated, but none of them are phenomenologically viable. 

 Finally,  we have identified a set of six mass matrices with no vanishing entries, and we have singled out  a unique case   
 that,  for normal ordering,  is  in excellent agreement with all available neutrino data,  and  predicts 
  $\mu_0 \sim 10\,$meV  for the lightest  neutrino mass,   $\delta \sim \frac{3\pi}{2}$  for the CP violating Dirac phase, 
 and it also determines the two Majorana phases via   the analytic expressions  
 in the second line of \eqn{eq:majoranaphasesOmega} (thus the rate of $0\nu2\beta$ decay can also be predicted,
 but it remains well below the sensitivity of present and foreseeable future experiments). 
We stress that for mass matrices of this type it would be impossible to identify the underlying $U(1)$ symmetry, 
 even in the limit of infinite experimental precision in the determination of all the low-energy parameters. This not only because 
 the symmetry leaves no obvious trace in the structure of the neutrino mass matrix, but  even more importantly because different symmetries 
 (in the sense of different charge assignments for the fields that yield different high-energy structures) can give rise to the same low-energy structure. For example,  
 the successful no-zero texture  that we have identified as viable and predictive, can be generated by the two different high-energy 
 structures that we have labeled as $\UVX{G}$:~$ 2 (2) [12][13]$   and  $\UVX{D}$:~$ 1 (3) [\,]\{13\}$.
 It is for this reason that it is appropriate to define this type of symmetries  as {\em covert}. 
 Additional information that could help discriminating among the different symmetries might come from 
 the study of leptogenesis, since this could give some insight into the structure of the fundamental 
 high-energy mass matrices.  For example,  
 in  $\UVX{G}$:~$ 2 (2) [12][13]$  the structure of  $M_N$ implies two degenerate states with opposite CP 
transformations which correspond to one heavy Dirac neutrino, and this strongly disfavours the possibility 
that leptogenesis could be realised in this scenario, while for  $\UVX{D}$:~$ 1 (3) [\,]\{13\}$ 
there is no apparent reason why  leptogenesis should not be successful.
A study of the high-energy signatures that could distinguish between $\UVX{G}$ and  $\UVX{D}$ is, however, 
beyond the scope of the present paper. 
 
One should also keep in mind that a fundamental ingredient of our construction  is that 
in general the lepton fields all carry different $U(1)$ charges.
Since $U(1)$ is spontaneously broken, there is a Nambu-Goldstone boson (NGB) $a(x)$  that couples non-diagonally to the lepton fields,  
so that lepton flavour violating (LFV) processes like $\mu \to e a$, $\tau \to \mu a$ etc.~will unavoidably occur.  
However, since all the couplings of this NGB are suppressed as $1/f_\phi$, if the value of this scale is sufficiently large, 
say  $f_\phi \gg 10^9\,$GeV,   the rate of all these processes can remain safely below the experimental bounds. 

There are other two issues that deserve some comments. 
After $U(1)$ is broken,  the  vanishing entries in the high energy matrices can get lifted 
by corrections, and it is  important to see if their size can be kept sufficiently small so that 
 the $f(m_{ij})\simeq 0$ conditions remain satisfied to a sufficiently good level of approximation.
The second issue is related to the presence of  two Higgs doublets coupled 
to the lepton fields, that can give rise to LFV processes that  must  also remain under control. 
Both these issues depend  on the details of the scalar sector. After $U(1)$ breaking 
the leading corrections to the  texture zero  will involve  the Higgs mixing term originating from the couplings in  \eqn{eq:h12phi}.
The relevant parameter  in case~1 is  $m^2_{12}= \mu  f_\phi$ with $\mu$ a coupling with dimension of a mass,  
 and in case~2  is $m^2_{12}= y  f_\phi^2 $ with $y$ some dimensionless coupling. 
To discuss one example, let us  consider the  textures of case $\mathbf{G}$, see  eq.~(\ref{eq:texturesG}).
Since the  Yukawa entry  in the Dirac matrix $(m_D)_{1,2}$ is non vanishing,  
from the couplings  $\bar N_1  \ell_2 H_1$   and  $\bar N_2  \ell_2 H_2$ 
one can draw a mixed self-energy loop diagram connecting $N_1$ and $N_2$.  
This requires  an insertion of the $U(1)$ breaking  spurion $m_{12}^2$ on the scalar field line and, 
recalling that two non-vanishing entries in the same row or column in $m_D$  must 
necessarily involve different Higgs doublets, this is indeed a general requirement.   Therefore, besides the suppression from 
 loop factors, it would  be also desirable to keep the size of $m^2_{12}$
as small as possible.   On the other hand, keeping under control Higgs mediated LFV interactions 
requires that below some sufficiently large scale one heavy Higgs   with mass $m_h$ will decouple, leaving 
only one light Higgs with  mass $m_l \sim 125\,$GeV  and  with the same properties of the SM Higgs. 
The two issues can be simultaneously solved for example by taking 
 in the scalar potential a large (positive) mass square parameter   $m^2_2 \sim m^2_h \gg v^2$ for, say, $H_2$, and a 
 relatively  smaller mixing term, e.g.  $m^2_{12}/m^2_h \sim 10^{-3}$. 
With $m_2 \sim 100\,$TeV all  LFV effects remain well under control, and  the size of the mixed $N_1$-$N_2$ 
self-energy diagrams, which is of order $\sim M_N \frac{Y_D^2}{16 \pi^2}  \frac{m_{12}^2}{m^2_h} $ with $Y_D$  
a Yukawa coupling,   gets an additional strong suppression $\sim 10^{-3}$. A scalar potential with these features might even help
to account for the lightness of the leptons coupled to $H_2$, since their couplings to the light Higgs 
that acquires a vev is suppressed by the mixing.

Finally, it could be tempting to interpret the $U(1)$ symmetry as a Peccei-Quinn (PQ) symmetry,  and $a(x)$ as the axion. In order to do so it is sufficient
to extend the symmetry to the quark sector. If the requirement of enforcing minimality in the number of parameters 
is maintained also for the quarks, then all the viable possibilities have already been classified in   
Ref.~\cite{Bjorkeroth:2018ipq}. In all cases the $U(1)$ symmetry has a QCD anomaly and thus it is a PQ symmetry.  In an extended model of this type, 
the high-energy scale $f_\phi$  would then acquire an even more fundamental role, 
since a value $f_\phi \sim 10^{10}-10^{12}\,$GeV would at the same time be a  
natural  one for the type-I seesaw, optimal for successful leptogenesis, safe with respect to FCNC and LFV decays, 
and preferred by the vast majority of axion models.

\section*{Acknowledgements}

F.B.~and E.N.~are supported by the INFN ``Iniziativa
Specifica'' Theoretical Astroparticle Physics (TAsP-LNF).  
The work of L.D.L.~is supported by the Marie Sk{\l}odowska-Curie Individual Fellowship grant AXIONRUSH (GA 840791) and the ERC grant NEO-NAT.
F.M.~is supported by MINECO grant FPA2016-76005-C2-1-P, by 
Maria de Maetzu program grant MDM-2014-0367 of ICCUB and 
2017 SGR 929.  
F.M.~acknowledges the INFN Laboratori Nazionali di Frascati 
for hospitality and financial support.

 \appendix

\section{ U(1)
charges for viable neutrino mass textures} 
\label{sec:viablecharges} 

We list in this Appendix the values of the $U(1)$ charges that yield the two viable textures for the neutrino mass matrix.  
The first possibility  corresponds to a structure of type $\UVX{E}$, and to the specific texture in which the neutrino mass 
matrix $m$  has two  vanishing entries in the $(1,1)$ and $(1,3)$  positions:
\begin{equation}
\label{eq:E}
 \UVX{E}: \quad  m = \ppmatr{0 & \times & 0 \\ \times & \times & \times \\ 0 & \times & \times}\,.
\end{equation}
There is  a unique form of the  matrices $M_N,\, m_D$ and $m_e$ which yields this low-energy texture which,
according to the notation introduced in Section~\ref{sec:structures},  is labeled as $2(3)[13][\,]$:
\begin{equation}
  \UVX{E}: \quad   M_N = 
    \ppmatr{0 & \times & \times \\ \times & 0 & 0 \\ \times & 0 & \times}
    , \quad
    m_D = 
      \ppmatr{ \times & 0 & \times  \\ 0 & \times  & 0 \\ 0 & 0 & \times}
    , \quad
    m_e =    
       \ppmatr{  \times & 0 & 0\\ 0 & \times  & 0 \\ 0 & 0 & \times}.
\end{equation}
Modulo an overall change of sign, the set of $U(1)$ charges for this case is also unique:
\begin{equation}
   \UVX{E}: \quad     \mX(N) = ( -3,1, 5) ,\quad
    \mX(\ell) = ( -5, -1,7) ,\quad
    \mX(e) = ( -7,1, 5),
\end{equation}
Numerical confrontations with up to date neutrino data have been recently carried out in 
Refs.~\cite{MadanSingh2019,Alcaide:2018vni}, and confirm that 
the two-zero texture mass matrix in \eqn{eq:E} is viable and predictive. 
The other viable and predictive possibility corresponds  to a mass matrix 
with no  zero textures, 
but with the entries constrained by the conditions in \eqn{eq:nozerorel} with   $[ij\bar{k}] = [132]$. 
Such a  texture can be generated by two different high-energy structures. One is  of type $\UVX{G}$  with label  $2(2)[12][13]$, 
as given in Table~\ref{tab:textures}. The mass matrices are: 
\begin{equation}
\label{eq:texturesG}
   \UVX{G}:\quad    M_N = 
    \ppmatr{ \times & 0 & 0 \\  0 & 0 &\times \\ 0 & \times & 0 }
    , \quad
    m_D = 
      \ppmatr{ \times &  \times & 0 \\ 0 & \times  & 0 \\ 0 & 0 & \times}
    , \quad
    m_e =    
       \ppmatr{  \times & 0 & \times \\ 0 &  \times & 0  \\ 0 & 0 & \times}, 
\end{equation}
which result from  a unique set of charges:
\begin{equation}
\label{eq:G}
 \UVX{G}:\quad   \mX(N) = ( 1,5,-7) ,\quad
    \mX(\ell) = ( -1, 3,-5) ,\quad
    \mX(e) = ( -3, 5-7)\,.   
\end{equation}
 The other is a structure of type $\UVX{D}$  with texture $1(3)[\,]\{13\}$, that is a permutation of the case listed in 
 Table~\ref{tab:textures}, and  for which the mass matrices are: 
\begin{equation}
 \UVX{D}:\quad  
    M_N = 
    \ppmatr{ \times & 0 & \times \\  0 & 0 &\times \\ \times & \times & 0 }
    , \quad
    m_D = 
      \ppmatr{ \times &  0 & 0 \\ 0 & \times  & 0 \\ 0 & 0 & \times}
    , \quad
    m_e =    
       \ppmatr{  \times & 0 & \times \\ 0 &  \times & 0  \\ 0 & 0 & \times} \ \mathrm{or} \ 
              \ppmatr{  \times & 0 & 0 \\ 0 &  \times & 0  \\ \times & 0 & \times}. 
\end{equation}
The set of charges that enforce these textures  can be written as: 
\begin{equation}
\label{eq:D}
 \UVX{D}:\quad  
  \mX(N) = ( 1,5,-3) ,\quad
    \mX(\ell) = ( 0, p,-2) ,\quad
    \mX(e) = ( q,p\pm1, q-2),  
\end{equation}
with $p=5\pm 1$ and $q=\pm1$. The structure  of these matrices is invariant with respect to 
a particular choice for $p$, while the different values of $q$  simply give $(m_e)_{13} \neq 0$ or $(m_e)_{31} \neq 0$.

\section{Signature of the eigenvalues    
for texture {\bf D} }  
\label{sec:signature} 

The structure of the neutrino mass matrix $\tilde m$ for texture $\UVX{D}$ in the 
 basis in which the CL charges are well defined is:
\begin{equation}
	\tilde m= \pmatr{m_{11} & 0 & \sqrt{m_{11}m_{33}} \\ 0 & 0 & m_{23} \\ \sqrt{m_{11}m_{33}} & m_{23}& m_{33}}. 
\label{eq:signD}
\end{equation}
We want to prove that, in the basis in which all the entries in $\tilde m$ are real, 
when ordered according to   increasing absolute values, 
the sign of the second eigenvalue $\mu_2$ is discordant from the signs of the 
 eigenvalues  with the smallest and largest absolute values $\mu_1$ and $\mu_3$.
Let us write down the coefficients of the characteristic equation  $\det(\tilde m-\mu I)=0$:
\begin{eqnarray}
\tr({\tilde m}) &=& m_{11} + m_{33} = \mu_1+\mu_2+\mu_3\,,  \\
 c_1 ({\tilde m}) &=& m_{23}^2 = -(\mu_1\mu_2+\mu_1\mu_3+ \mu_2\mu_3)\,, \\
\det({\tilde m} )&=& - m_{11} m_{23}^2 =\mu_1\mu_2\mu_3 \, ,
\end{eqnarray}
where $c_1$ is the coefficient of the linear term. From this we can derive a   useful relation:
\begin{eqnarray}
 R &=& -\left[c_1 (\tilde m) \tr(\tilde m) + \det(\tilde m )\right] = - m_{33} \, m_{23}^2 =(\mu_1+\mu_2)(\mu_1+\mu_3)(\mu_2+\mu_3)\,. 
\end{eqnarray}
Now the signs of $m_{11}$ and $m_{33}$ must be concordant to ensure that $m_{13}$ is real. 
If   $m_{11},m_{33} >0$  the determinant is negative and we have either   three negative eigenvalues 
or one negative and two positive. Since $\tr(\tilde m)>0$ the first possibility is excluded. 
Next we have to ensure that $R <0$.  It is readily seen that the ordered signatures $(-,+,+)$
and $(+,+,-)$  for the eigenvalues give $R >0$ and are excluded, while $(+,-,+)$ ensures the correct result. If $m_{11},m_{33} <0$
the same reasoning gives the signature  $(-,+,-)$, hence the  sign of $\mu_2$ is always discordant from 
that of $\mu_1$ and $\mu_3$. The two cases $m_{11},m_{33} >0$ and $m_{11},m_{33} <0$
are related by a trivial rephasing of all the lepton fields by $e^{i\pi/2}$. Therefore assuming 
the eigenvalues signature $(+,-,+)$ is without loss of generality.

\bibliographystyle{JHEP}




\providecommand{\href}[2]{#2}\begingroup\raggedright\endgroup

\end{document}